\documentclass[twoside,twocolumn,english,aps,showpacs,prl,superscriptaddress]{revtex4-1}

	\usepackage[usenames,dvipsnames]{xcolor}
	\usepackage{amsmath}
	\usepackage{amsfonts}
	\usepackage{amssymb}
	\usepackage[colorlinks=true,citecolor=blue,linkcolor=red]{hyperref}
	\usepackage{graphicx}
	\usepackage{bbold}					
	\usepackage[makeroom]{cancel}		
	\usepackage{multirow}				
	\usepackage[normalem]{ulem}        
	\usepackage{array}
	\usepackage{comment}
	\usepackage{makecell}
	\usepackage{pdfpages}
	\makeatletter
	\AtBeginDocument{\let\LS@rot\@undefined}
	\makeatother

\makeatletter

\makeatletter
\newcommand*{\rom}[1]{\expandafter\@slowromancap\romannumeral #1@}
\makeatother

\usepackage{times}{\Large }

\makeatother

\usepackage{babel}
\begin{document}

\title{Universal conductance scaling of Andreev reflections using a dissipative probe}

\author{Donghao Liu}
\thanks{These authors contributed equally to this work.}
\affiliation{State Key Laboratory of Low Dimensional Quantum Physics, Department of Physics, Tsinghua University, Beijing, 100084, China}

\author{Gu Zhang}
\thanks{These authors contributed equally to this work.}
\affiliation{Beijing Academy of Quantum Information Sciences, Beijing 100193, China}

\author{Zhan Cao}
\affiliation{Beijing Academy of Quantum Information Sciences, Beijing 100193, China}

\author{Hao Zhang}
\affiliation{State Key Laboratory of Low Dimensional Quantum Physics, Department of Physics, Tsinghua University, Beijing, 100084, China}
\affiliation{Beijing Academy of Quantum Information Sciences, Beijing 100193, China}
\affiliation{Frontier Science Center for Quantum Information, Beijing 100184, China}

\author{Dong E. Liu}
\email{Corresponding to: dongeliu@mail.tsinghua.edu.cn}
\affiliation{State Key Laboratory of Low Dimensional Quantum Physics, Department of Physics, Tsinghua University, Beijing, 100084, China}
\affiliation{Beijing Academy of Quantum Information Sciences, Beijing 100193, China}
\affiliation{Frontier Science Center for Quantum Information, Beijing 100184, China}

\begin{abstract}
The Majorana search is caught up in an extensive debate about the false-positive signals from non-topological Andreev bound states (ABSs).  We introduce a remedy using the dissipative probe to generate electron-boson interaction. We theoretically show that the interaction-induced renormalization leads to significantly distinct universal zero-bias conductance behaviors, i.e. distinct characteristic power-law in temperature, for different types of Andreev reflections, which shows a sharp contrast to that of a Majorana zero mode. Various specific cases have been studied, including the cases that two charges involved in an Andreev reflection process maintain/lose coherence, and the cases for multiple ABSs with or without a Majorana present. A transparent list of conductance features in each case is provided to help distinguishing the observed subgap states in experiments, which also promotes the identification of Majorana zero modes.
\end{abstract}


\date{\today}

\maketitle


\textbf{{\em Introduction.}} Quantum tunneling~\cite{wolf2012principles} has been used as a very powerful method to study quantum materials and quantum devices.
However, if obtained using non-interacting probe and target, the tunneling signal is usually very sensitive to contaminants that potentially induce non-universal behaviors.
As an important example, the tunneling spectroscopy signals in detecting Majorana zero modes (MZMs)~ \cite{ReadGreen,1DwiresKitaev} in semiconductor-superconductor heterostructures~\cite{LutchynPRL10,oreg2010helical,lutchyn2018majorana} should give a quantized zero-bias conductance peak~\cite{sengupta2001midgap,Law09,flensberg2010tunneling,wimmer2011quantum} and a robust quantized plateau by varying all relevant control parameters. However, the current experimental results~\cite{Mourik2012,deng2012anomalous,das2012zero,finck2013anomalous,churchill2013superconductor,deng2016majorana,ZhangNC2017,chen2017experimental,suominen2017zero,nichele2017scaling,ZhangNN2018,vaitiekenas2018effective,de2018electric,bommer2019spin,grivnin2019concomitant,pan2020arXiv,zhang2021large,song2021large} are far from the ideal predictions. One of the key reasons is that such a non-interacting Majorana detection platform is easily contaminated by junction and disorder-induced Andreev bound states (ABSs)~\cite{pientka2012enhanced,liu2012zero,cole2016proximity,CXLiuABSMZM,liu2018impurity,cao2019decay,pan2020physical,sarma2021disorder,pan2021quantized} that cause non-robust signals.

As a possible remedy, interaction, known as a method to sharpen transition between different fixed points, can be introduced to classify different physics under the interaction renormalization~\cite{cardy1996scaling}.
Indeed, different physics emerges near fixed points that belong to distinct interaction-dependent universality classes.
One of the simplest schemes to introduce interaction is to consider a dissipative electromagnetic environment, e.g., applying an ohmic resistance in series with the tunneling junction, and causing effective electron-boson interaction~\cite{Nazarov1992charge}. With ohmic dissipation, the tunneling conductance exhibits dissipation-dependent power-law scaling behaviors~\cite{Nazarov1992charge}.
Various kinds of dissipation-influenced charge transport in mesoscopic systems have been studied both experimentally~\cite{Dong2012Nature,MebrahtuNaturePhy,JezouinPierre13,AnthorePierrePRX18} and theoretically~\cite{MatveevGlazman93,SafiSaleurPRL04,LeHurLiPRB05,FlorensPRB07,liu2013filter,DongPRB2014,HuaixiuPRB14,FlensbergDissipation2015,GuPRL17,LeHurDrivenQImpReviewCRP18,LDH2020PRB,CrossoverPRB21}.

\begin{figure}[t]
\includegraphics[width=1\columnwidth]{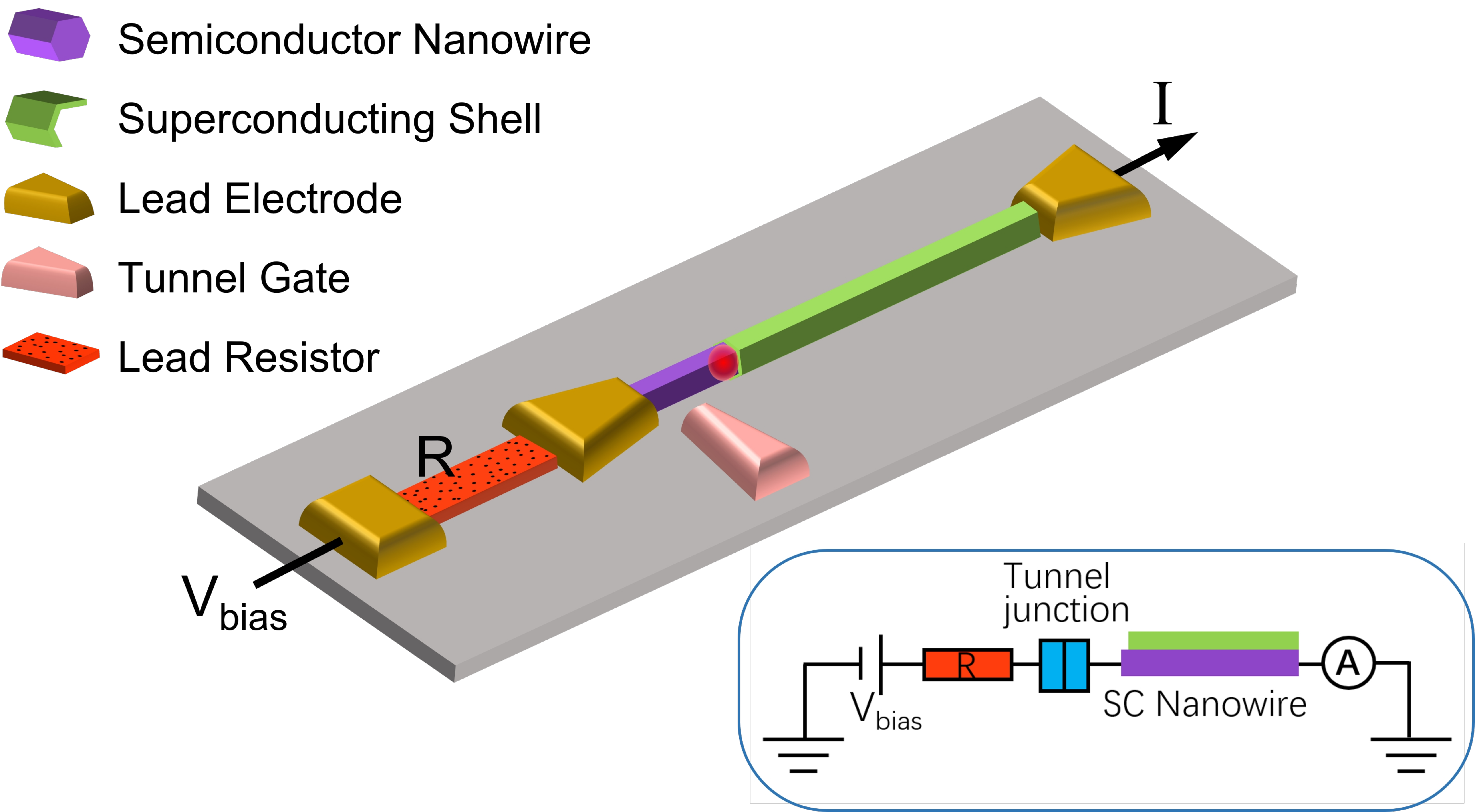}
\vspace*{-0.3cm}
\caption{Illustration of the system setup. The inset is the equivalent circuit. A bias voltage $V_\text{bias}$ is applied on the lead.}
\label{fig:setup}
\vspace*{-0.3cm}
\end{figure}

\textbf{{\em Main results.}} Here, we focus on the hybrid semiconductor superconductor nanowire device for the detection of Majorana resonance. The dissipative tunneling into a MZM was proposed as a Majorana signature filter~\cite{liu2013filter} by one of the authors. It is believed that junction and/or disorder-induced fermionic ABSs dominate the phase diagram of the current nanowire devices~\cite{sarma2021disorder,zeng2021partially}. Therefore, it is an interesting topic to study the dissipative interacting probe for different types of ABSs and obtain their universal scaling behaviors under renormalization to distinguish between ABSs and the real MZM.
Following this motivation, we map the dissipative tunneling into ABSs to a Coulomb gas model, and find that the zero-bias conductance has special power-law dependence on temperature $T$. In addition, we find that the coherence between electron and hole after the Andreev reflection will significantly affect the power-law behaviors; and therefore, the measurement of the power-law could be applied to detect the coherence after the Andreev reflection. We show that the dissipation can cause significantly different universal behaviors according to their characteristic power law as summarized in Table~\ref{tab:Scaling} for different types Andreev reflections and MZM, where $r = Re^2/h$ is the dimensionless dissipation amplitude. Our result also potentially applies to other platforms using scanning tunneling microscope ~\cite{JJinFeng-2016-PRL,DHong-2018-Sci,FDongLai-2018-PRX,KLingYuan-2019-NatPhys,Hanaguri-2019-NatM}, where dissipation-induced dynamical Coulomb blockade features have been observed~\cite{STM-DCB-2012PRL}.

\begin{table}[ht]
\vspace*{-0.4cm}
\caption{\label{tab:Scaling} Universal Scaling Behaviors of different circumstances.}
\begin{tabular}{|c|c|c|c|c|}
\hline 
\multicolumn{1}{|c|}{} & Majorana~\cite{liu2013filter} & \makecell{ABS \\ (coherent)} & \makecell{ABS \\ (incoherent)} & \makecell{Normal\\ state~\cite{Nazarov1992charge}}\tabularnewline
\hline 
\hline 
\multicolumn{1}{|c|}{Qualitative} & Enhancement & Decay & Decay & Decay\tabularnewline
\hline 
\multicolumn{1}{|c|}{\makecell{Universal\\ scaling}} & $\frac{2e^{2}}{h}-G\sim T^{\frac{2-4r}{1+2r}}$ & $G\sim T^{8r}$ & $G\sim T^{4r}$ & $G\sim T^{2r}$\tabularnewline
\hline 
\end{tabular}

\end{table}

\textbf{\em{Model.}} An illustration of the system is shown in Fig.~\ref{fig:setup}. The dissipative tunneling to an ABS is achieved by tunnel coupling
a lead to a nanowire. We consider the hybrid semiconductor nanowire$-$superconductor devices with finite magnetic field~\cite{LutchynPRL10,oreg2010helical}. This type of devices are recently well-studied for the detection of MZMs. Later we also call this hybrid device the ``SC nanowire''. In reality, devices of this type are easily contaminated by disorders in the junction or the nanowire bulk, and the disorder-induced trivial ABSs potentially produce false-positive signals in the standard tunneling experiments. The system is also coupled to a dissipative bath, which can be achieved by replacing part of the electrode with a thin long resistive metal strip {[}with resistance $R$; red in Fig.~\ref{fig:setup}(a){]}. The tunnel-gate controls the coupling between the lead and the SC nanowire. An equivalent circuit diagram is shown in Fig.~\ref{fig:setup}(b).





\begin{figure}
\includegraphics[width=1\columnwidth]{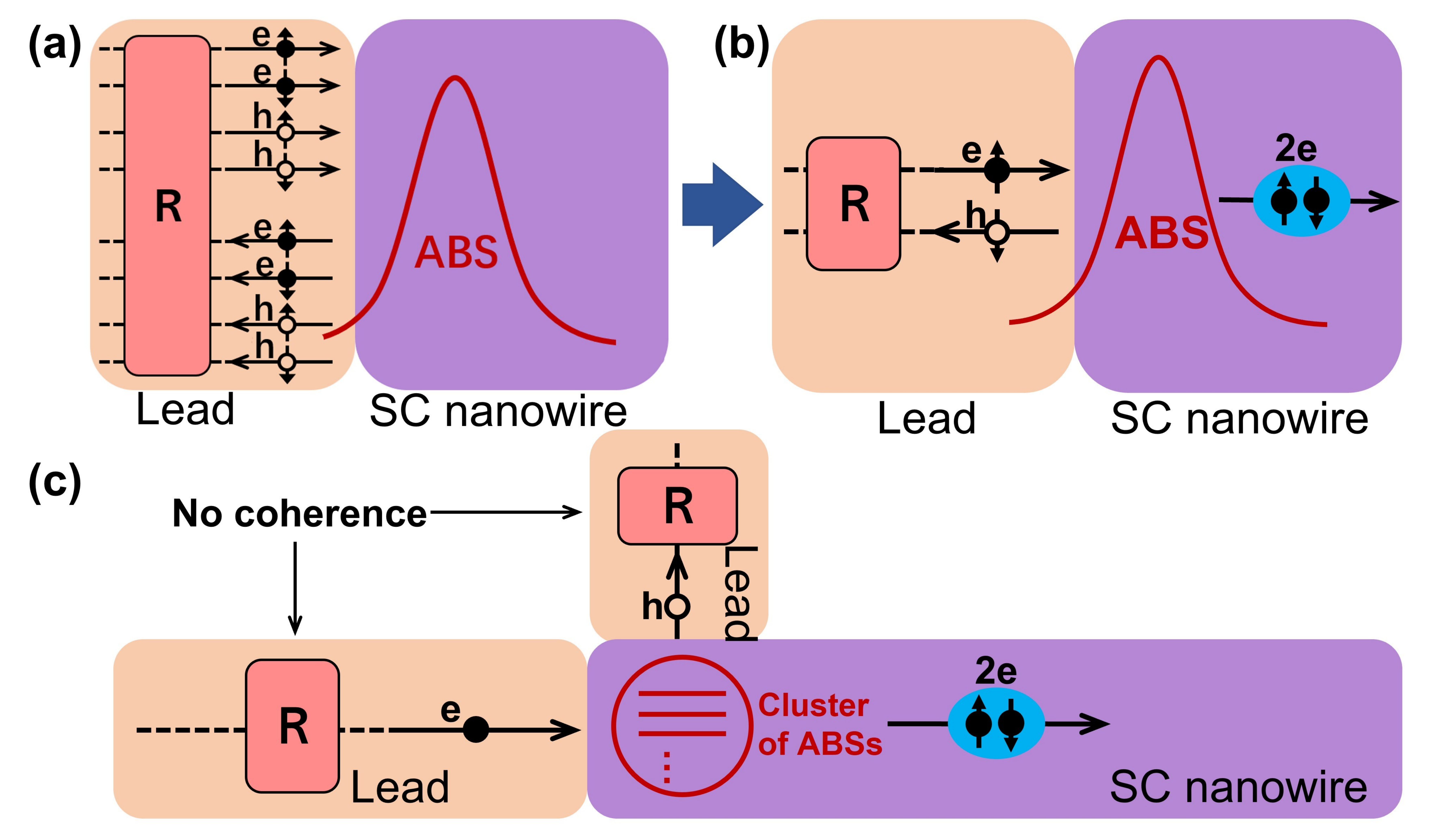}\caption{\label{fig:AllChannels}(a) All possible processes.
 (b) Leading process after RG. (c)Incoherent dissipative Andreev reflection via a cluster of ABSs.}
\end{figure}

The whole system includes four parts: the SC nanowire, the lead, the tunneling part, and the dissipative environment. For the SC nanowire, we focus on the case with an ABS localized at the left side of the nanowire, and its Hamiltonian can be written as $H_{\text{wire}}=\epsilon a^{\dagger}a+\text{const,}$ where $a$ is the fermionic quasiparticle operator for the ABS. 
Its energy is inside the superconducting gap $\epsilon<\Delta$ and can reach $\epsilon\rightarrow0$ by adjusting experimental variables. The lead can be described by the Hamiltonian of spinful fermions with
dispersion linearized close to the Fermi energy:
\begin{equation}
H_{\text{lead }}=v_{F}\!\sum_{\sigma=\uparrow/\downarrow}^{N}\!\int_{-\infty}^{0}\!dx \,\psi_{L,\sigma}^{\dagger}i\partial_{x}\psi_{L,\sigma}-\psi_{R,\sigma}^{\dagger}i\partial_{x}\psi_{R,\sigma},\label{leadH}
\end{equation}
where $\psi_{L,\sigma}\left(x\right)$ and $\psi_{R,\sigma}\left(x\right)$
are fermion operators for the left-moving and right-moving modes with
spin $\sigma$ at point $x$ in the lead. 
Counting the degrees of freedom of the particle (hole) and the spin, there are a total of four conducting channels in the lead. The tunneling part of the Hamiltonian can be expressed as
\begin{equation}
H_{T}=\left(\hspace{-0.0em}
\psi_{\uparrow}^{\dagger}\left(0\right) \hspace{+0.2em} \psi_{\downarrow}^{\dagger}\left(0\right)\hspace{-0.0em}\right)\left(\hspace{-0.3em}\begin{array}{cc}
t_{e\uparrow} & t_{h\uparrow}^{*}\\
t_{e\downarrow}^{*} & t_{h\downarrow}
\end{array}\hspace{-0.3em}\right)\left(\hspace{-0.3em}\begin{array}{c}
a\\
a^{\dagger}
\end{array}\hspace{-0.3em}\right)e^{-i\varphi}+h.c.,
\label{eq:TunnelHamiltonian}
\end{equation}
where $\psi_{\sigma}(0) = \psi_{L,\sigma}(0) + \psi_{R,\sigma}(0)$.
Breaking the spin rotation and time reversal symmetries, we need four independent tunneling parameters $t_{e\uparrow}$, $t_{e\downarrow}$, $t_{h\uparrow}$, $t_{h\downarrow}$ to describe an arbitrary ABS, as shown in Fig.~\ref{fig:AllChannels}(a).
The operator $e^{-i\varphi}$ is conjugate to the charge fluctuation $Q$ of the junction capacitance, following $\left[\varphi,Q\right]=ie$, and thus accompanies nanowire-superconductor charge transport.
It couples bilinearly to the dissipative environment represented by a set of harmonic oscillators (i.e., $\left\{ q_{n},\varphi_{n}\right\} $ with oscillator frequency $\omega_{n}=1/\sqrt{L_{n}C_{n}}$)~\cite{leggett1987dynamics,Nazarov1992charge,caldeira1981influence}:
$H_{\text{env}}={Q^{2}}/{2C}+\sum_{n=1}^{N}\left[{q_{n}^{2}}/{2C_{n}}+\left({\hbar}/{e^{2}}\right)^{2}\left(\varphi-\varphi_{n}\right)^{2}/{2L_{n}}\right]$,
where $C_n$ and $L_n$ respectively refer to the effective capacitance and impedance of the $n$th dissipative mode.

\textbf{\em{Effective action and Coulomb gas model.}} Overall, the  partition function of this tunnel junction coupled to the dissipative environment is
\begin{equation}
Z=\int \left[D\Phi_{\uparrow}\right]\left[D\Phi_{\downarrow}\right]\left[D\varphi\right] \left[Da\right] e^{-S_{\text{eff}}}e^{-S_{\text{T}}},\label{eq:partition}
\end{equation}
where $S_{\text{T}}$ is the action of the tunneling part $S_{\text{T}}=\int_{0}^{\beta}d\tau L_{\text{T}}$  with the tunneling Lagrangian $L_{\text{T}}$ directly obtained from Eq.~\eqref{eq:TunnelHamiltonian}, and $\beta$ is the temperature inverse.
After a spatial integral, the effective action becomes $S^{\text{eff}}\!\!=\!\frac{1}{\beta}\sum_{\omega_n}\left|\omega_{n}\right|\left[\sum_{\sigma}\left|\Phi_{\sigma}\left(\omega_{n}\right)\right|^{2}+\left|\varphi \left(\omega_{n}\right) \right|^{2}/2r\right]$, with the first and the second terms in the brackets from the lead and the environment parts respectively. 
For later convenience, we define $r = Re^2/h$ as the dimensionless dissipation.
 $\Phi_{\sigma}\left(x\right)$ is the chiral Bosonic field from the standard Bosonization~\cite{Giamarchi:2003}:  
 $\psi_{L/R,\sigma}\left(x\right)=\frac{1}{\sqrt{2\pi\alpha}}F_{\sigma}e^{-i\Phi_{\sigma}\left(\mp x\right)}$,
where $\alpha$ is the short-distance cutoff and $F_{\sigma}$ is
the Klein factor.
In the action $S^{\text{eff}}$, the lead part is obtained by integrating out fluctuations in $\Phi_{\sigma}\left(x\right)$ away from $x=0$~\cite{anderson1970exact,KaneFisher2,KaneFisher3,furusaki1993resonant,SupMat}, and the environment is obtained by integrating out the environmental degree of freedom~\cite{WeissBook}.

By expending the partition function Eq. (\ref{eq:partition}) in the powers of tunneling part and then integrating out the bosonic field, we can obtain the Coulomb gas representation~\cite{anderson1970exact,KaneFisher2} for our model
\begin{align}
Z=&\sum_{\nu=\pm1}\hspace{-0.1em}\sum_{n}\hspace{-0.1em}\sum_{\left\{\hspace{-0.1em} q_{i}\hspace{-0.1em}\right\} }\hspace{-0.1em}\sum_{\left\{ \hspace{-0.1em} s_{i} \hspace{-0.1em}\right\} }\hspace{-0.1em}C_{\text{t}}\hspace{-0.1em}\int_{0}^{\beta}\hspace{-0.3em}d\tau_{2n}\hspace{-0.3em}\int_{0}^{\tau_{2n}}\hspace{-0.3em}d\tau_{2n-1}\hspace{-0.1em}\nonumber\\
&...\hspace{-0.1em}\int_{0}^{\tau_{3}}\hspace{-0.3em}d\tau_{2}\hspace{-0.3em}\int_{0}^{\tau_{2}}\hspace{-0em}d\tau_{1}\hspace{-0em}e^{\sum_{i>j}\hspace{-0.2em}V_{ij}}\hspace{-0em} e^{ \nu\epsilon\left[\frac{\beta}{2}+\sum_{i}\left(-1\right)^{i}\tau_{i}\right]} ,\label{eq:CGModel}
\end{align}
which describes a 1D plasma of logarithmically interacting charges. The interaction $V_{ij}$ has the form:
\begin{align}
V_{ij}=\frac{1}{2g}\biggl[ &q_{i}q_{j}+K_{1}\left(q_{i}r_{j}+r_{i}q_{j}\right)+g s_{i}s_{j}\nonumber\\
&+K_{2}g\left(s_{i}r_{j}+r_{i}s_{j}\right)\biggr]
\text{ln}\left(\frac{\tau_{i}-\tau_{j}}{\tau_{c}}\right),\label{eq:VijGeneral}
\end{align}
where the effective interaction parameter $g=\left(1+4Re^2/h\right)^{-1} = (1 + 4r)^{-1}$, and $1/\tau_c$ refers to the high-energy cutoff that changes during each RG step.
Three types of charges, i.e., $q_{i}$, $s_{i}$ and $r_{i}$ are involved.
The first two  refer to the changes in charge and spin in the lead. The last one is from the ABS state.
Initially, $K_1 = K_2 = 0$.
They begin to grow in the presence of asymmetry, driving the system towards different fixed points~\cite{SupMat}.

\textbf{\em{RG analysis and scaling behaviors.}} In the framework of the Coulomb gas model, the renormalization group (RG) equation at weak tunneling coupling fixed point can be obtained from integrating out the degrees of freedom between $\tau_{c}$ and $\tau_{c}+d\tau_{c}$ (a real-space RG) \cite{SupMat}. The resulting RG equations yield
\begin{subequations}
\begin{align}
\frac{dK_{1}}{d\ln\tau_{c}}=&-2\tau_{c}^{2}
\left[\left(\left|t_{e\uparrow}\right|^{2}-\left|t_{h\uparrow}\right|^{2}+\left|t_{e\downarrow}\right|^{2}-\left|t_{h\downarrow}\right|^{2}\right)\right.\nonumber\\
&\left.+\left(\left|t_{e\uparrow}\right|^{2}+\left|t_{h\uparrow}\right|^{2}+\left|t_{e\downarrow}\right|^{2}+\left|t_{h\downarrow}\right|^{2}\right)K_{1}\right],
\label{eq:rg_k1}\\[3pt]
\frac{dK_{2}}{d\ln\tau_{c}}=&-2\tau_{c}^{2}\left[\left(\left|t_{e\uparrow}\right|^{2}-\left|t_{h\uparrow}\right|^{2}-\left|t_{e\downarrow}\right|^{2}+\left|t_{h\downarrow}\right|^{2}\right)\right.\nonumber\\
&\left.+\left(\left|t_{e\uparrow}\right|^{2}+\left|t_{h\uparrow}\right|^{2}+\left|t_{e\downarrow}\right|^{2}+\left|t_{h\downarrow}\right|^{2}\right)K_{2}\right],
\label{eq:rg_k2}\\[3pt]
\frac{dt_{\xi}}{d\ln\tau_{c}}= & \left[1-\frac{\left(K_{1}+\delta_{\xi,1}\right)^{2}+g\left(K_{2}+\delta_{\xi,2}\right)^{2}}{4g}\right]t_{\xi},\label{eq:rg_t1}\\[3pt]
\frac{d\epsilon}{d\ln\tau_{c}}= &
\ \epsilon,
\label{eq:rg_energy}
\end{align}
\label{eq:rg_general}
\end{subequations}
where $\xi$-dependent charge pair $\left(\delta_{\xi,1}, \delta_{\xi,2}\right)$ equals $\left(+1, +1\right)$, $\left(-1, -1\right)$, $\left(+1, -1\right)$, $\left(-1, +1\right)$ respectively, when $\xi$ $= e\hspace{-0.05em}{\uparrow}$, $h\hspace{-0.05em}{\uparrow}$, $e\hspace{-0.05em}{\downarrow}$ and $h\hspace{-0.05em}{\downarrow}$.
For simplicity, we begin with the fine-tuned situation where $\epsilon = 0$.
Initially, $K_1 = K_2 = 0$, and all tunneling operators of Eq.~\eqref{eq:TunnelHamiltonian} share the same scaling dimension [i.e., the factor after the minus sign of Eq.~\eqref{eq:rg_t1}] $1/2 + r$.
Following Eq.~\eqref{eq:rg_general}, the system flows to different fixed points depending on the symmetry among tunneling parameters $t_\xi$.

As the starting point, we look into the most generic situation and impose no requirement on tunneling parameters. Of this situation, absolute values of parameters $K_1$ and $K_2$ increase during the RG flow.
Accompanying their enhancement, scaling dimension of the tunneling with the strongest amplitude begins to decrease, which in term induces an even stronger asymmetry or difference among tunnelings $\propto t_\xi$.
To obtain a more intuitive understanding, we consider the fixed point where $t_{e\uparrow} \sim 1 \gg t_{e\downarrow}, t_{h\uparrow}$, $t_{h\downarrow}$, and $K_1 = K_2 = -1$. At this point, the leading process $\propto t_{e\uparrow}$ has a vanishing scaling dimension, and grows as if it was an energy cutoff.
Consequently, at low enough temperatures, $t_{e\uparrow} \psi^{\dagger}_\uparrow a + h.c.$ becomes infinite, where $\psi_\uparrow$ completely hybridizes the impurity ABS, and at meanwhile, suppresses the other communications between the ABS and the lead.
A persistent lead-superconductor transport now has to rely on coherent Andreev tunneling $t_{e\uparrow}t_{h\downarrow}\psi_{\uparrow}^{\dagger}(0)\psi_{\downarrow}^{\dagger}(0)e^{-2i\varphi} a^\dagger a + h.c.$.
As an Andreev tunneling consists of two coherent tunnelings [Fig.\,\ref{fig:AllChannels}(b)], its low-energy feature is determined by the less relevant process $\propto t_{h\downarrow}$.
Following Eq.~\eqref{eq:rg_general}, this process flows $dt_{h\downarrow}/d\ln \tau_c = -4 r t_{h\downarrow}$ when $K_1 = K_2 = -1$, with the scaling dimension $1 + 4r$.
This scaling dimension indicates that at low enough temperatures, the zero-bias conductance decreases following the temperature power-law $G\propto T^{8r}$ and vanishes at zero temperature, different from the regular dissipation tunneling ($G \propto T^{2r}$ ) without superconductivity.
As a reminder, the temperature power equals twice the difference between the scaling dimension and unity~\cite{KaneFisher3}.

This $8r$-power-law, however, requires perfect coherent Andreev reflections. In reality, the imperfection of the superconductor-proximitized nanowire induces possible transient states with the typical lifetime $t_\text{imp}$.
Even when two sub-processes of an Andreev tunneling are coherent, the relaxation of two involved charges, during which dissipation is produced, might lose coherence in an imperfect nanowire \cite{Incoherence}.
Indeed, we consider the Andreev tunneling operator $\mathcal{O}_\text{AR} =  \psi^{\dagger}_\uparrow a \exp(-i\varphi) \psi^\dagger_{\downarrow} a^\dagger \exp(-i\varphi) + h.c.$, whose correlation in time becomes
\begin{equation}
\begin{aligned}
  & \big\langle \mathcal{O}_\text{AR} (t) \mathcal{O}_\text{AR} (0) \big\rangle   = \big\langle H^0_\text{T}(t) H^0_\text{T}(0) \big\rangle  \\
   \cdot & \big\langle e^{-i\varphi(t + \delta t)} e^{-i\varphi(t + \delta t')} e^{i\varphi(\delta t)} e^{i\varphi(\delta t')} \big\rangle,
\end{aligned}
\label{eq:coherent_correlation}
\end{equation}
where $H^0_\text{T} = \psi^{\dagger}_\uparrow a \psi^\dagger_{\downarrow} a^\dagger$ contains the lead operators, and $\delta t,\ \delta t' \sim t_\text{imp}$ refer to the incoherence-induced delay in time~\cite{SupMat}.
Their amplitudes determine the leading feature of the correlation Eq.~\eqref{eq:coherent_correlation}. Indeed, when $t\gg t_\text{imp}$, the delay in time becomes negligible, where the correlation of two dissipative phases $\propto  t^{-8r}$. In contrast, the correlation changes to be $\propto t^{-4r}$ for relatively shorter time $t\ll t_\text{imp}$.
The correlation of phase is then determined by the cutoff in time, i.e., the inverse of the temperature $1/T$.

Extremely, when $t_\text{imp} \gg 1/T$, two phases become completely uncorrelated~\cite{SupMat}, thus reducing the suppression of conductance from dissipation by half [see Fig.\,\ref{fig:AllChannels}(c)]. In this limit, Andreev reflection has the scaling dimension $1 + 2r$ instead, leading to the conductance power-law $G \propto T^{4r}$.
As a possible extension, the coherence-dependent conductance power-law, which could potentially exist in other dissipative systems, provides us a possible tool in the detection of system coherence.
In our system, incoherent power-law might also occur in relatively high-temperature systems where ABS has not been fully hybridized by the dominant lead operator. In this situation, an incoming electron might stay on the ABS for a long-enough time during which the incoming electron and the reflected hole have become incoherent.

\textbf{\em{Conductance peaks.}} Experimentally, two types of conductance peaks might be observed. Firstly, and the most interestingly, a Majorana resonance peak emerges when normal lead couples either to a topological MZM, or when the ABS consists of two spatially separated MZMs that decouple from each other (i.e., the quasi-MZM scenario predicted by e.g.,~\cite{Aguado-QDABS2012,kells2012PRB,moore2018twoTerminal,moore2018quantized,quasiMajoranaWimmer2019}) in case of a smooth lead-wire barrier~\cite{kells2012PRB}.
These two situations are indistinguishable from local measurements, and luckily, both display non-trivial behaviors including non-Abelian statistics (see, e.g.,\cite{quasiMajoranaWimmer2019}).
In our model, an accidental MZM scenario occurs when fine-tuning $|t_{e\uparrow}|^2 - |t_{h\uparrow}|^2 + |t_{e\downarrow}|^2 -  |t_{h\downarrow}|^2 = 0$, and $t_{e\uparrow} t_{h\downarrow} = t^*_{e\downarrow} t^*_{h\uparrow}$, where only $(a + a^\dagger)/\sqrt{2}$ couples to the lead Majorana $(\psi_{\uparrow'} - \psi^{\dagger}_{\uparrow'})/\sqrt{2}$, where $\psi^\dagger_{\uparrow'}$
refers to the fermion with the spin along the direction determined by relative amplitudes of $t_{e\uparrow} + t^*_{h\uparrow}$ and $t^*_{e\downarrow} + t_{h\downarrow}$ (e.g., $\uparrow'$ points towards the $x$ direction if $t_{e\uparrow} + t^*_{h\uparrow} = t^*_{e\downarrow} + t_{h\downarrow}$).
Of this scenario, two asymmetry parameters $K_1 = K_2 = 0$ are fixed at zero, indicating the protection of the zero-bias conductance peak $2e^2/h$ by system symmetry at zero temperature.
Specifically, as the lead-Majorana coupling has the scaling dimension $r + 1/2$ initially, its dual operator at low temperatures has the scaling dimension $2/(1 + 2r)$, the inverse of the initial scaling. From that one arrives at the power law $2e^2/h-G\sim T^{\frac{2-4r}{1+2r}}$ that agrees with the result of Ref.\,\cite{liu2013filter}, where tunneling to a real Majorana is considered.

\begin{figure}
\includegraphics[width= 0.9 \columnwidth]{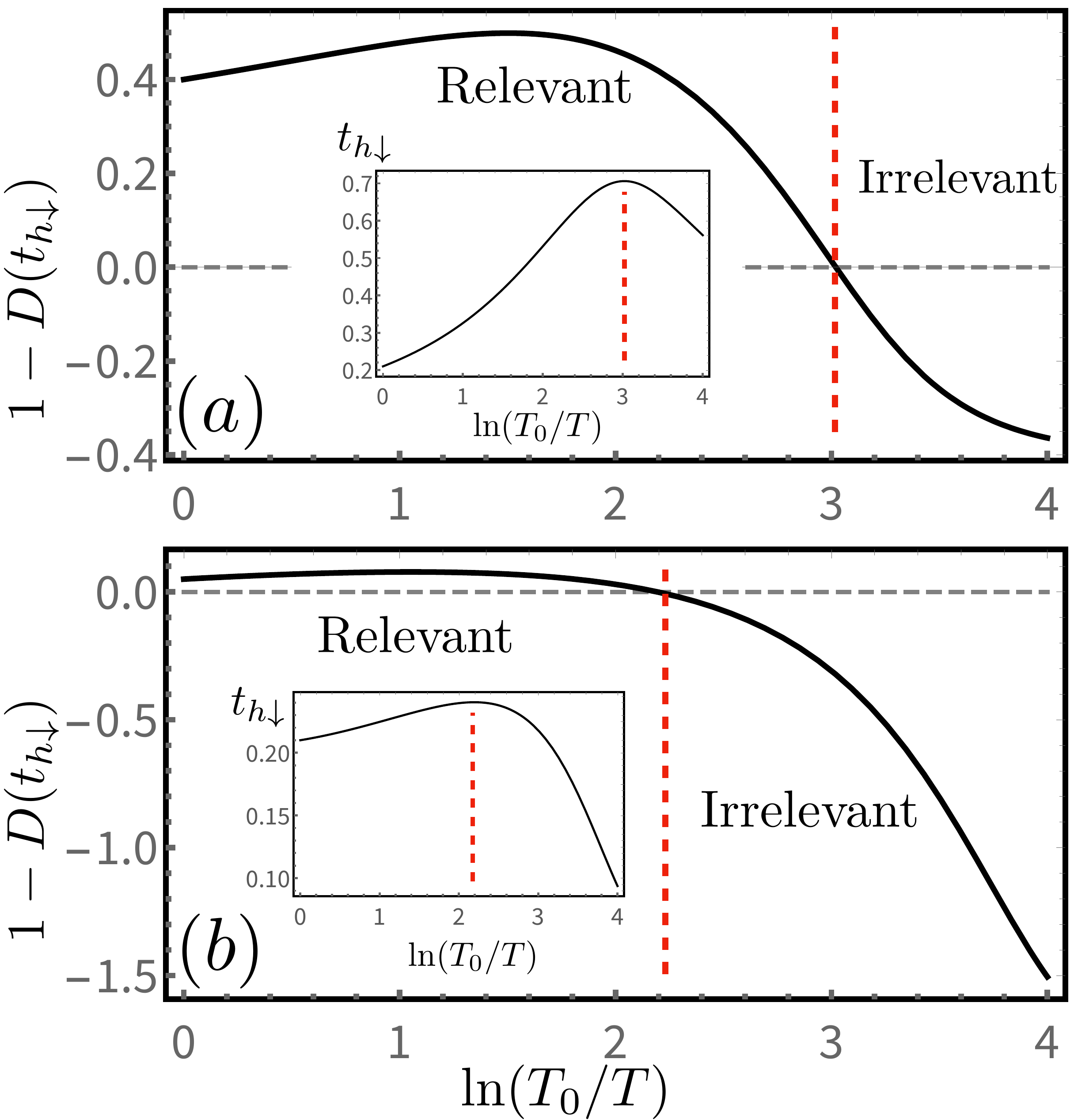}
\caption{Scaling dimension and amplitude (inset) of the operator $\propto t_{h\downarrow}$, when $t_{e\uparrow}$ is dominating.
Temperature decreases when value of $x$-axis increases.
(a) When $r = 0.1$, $t_{h\downarrow}$ becomes RG irrelevant when $T \approx 0.05 T_0$, with the peak height $\approx 0.7$. (b) When $r = 0.45$, the ABS conductance arrives at its peak position much faster $T \approx 0.11 T_0$, with a much smaller peak height $\approx 0.24$. Other parameters in (a) and (b) are the same.}
\label{fig:abs_peak}
\end{figure}

Although theoretically any dissipation is capable of killing generic (i.e., not fine-tuned) ABS peaks at low enough energies, in real experiments ABS peaks might emerge when electron temperature is too high to witness the conductance suppression from a weak dissipation.
This is especially true for a weak asymmetry among tunnelings parameters.
Nevertheless, we expect the absence of universality near these ABS peaks, as they do not correspond to fixed points in a dissipative system following RG equations \eqref{eq:rg_general}.
We emphasize that the missing of universality near the ABS peak does not contradict the materials of Table.\,\ref{tab:Scaling}, where scaling behaviors are only predicted near the zero-conductance fixed point, at low-enough temperatures.
For instance, we study the case where $t_{e\uparrow} = 0.25$ is slightly larger than $t_{h\downarrow}$ and much larger than the other two parameters.
In this situation the system conductance and its dependence on temperature are mostly determined by $t_{h\downarrow}$.
In Fig.\,\ref{fig:abs_peak}, we thus plot $t_{h\downarrow}$ and its scaling dimension $D(t_{h\downarrow})$ [i.e., $\left(K_{1}-1\right)^{2}/4g+\left(K_{2}+1\right)^{2}/4$ of Eq.~\eqref{eq:rg_t1}] to indirectly investigate the conductance features.
In Fig.\,\ref{fig:abs_peak}(a), for a weak dissipation $r = 0.1$, the conductance arrives at its peak value [where $D(t_{h\downarrow}) = 1$ ] when $T \approx 0.05 T_0$, where $T_0$ refers to temperature at which RG starts. However, near this peak position, $D(t_{h\downarrow})$ keeps changing with temperature, indicating the absence of universality, i.e., persuasive temperature power-laws of conductance under low-enough energies.
For a stronger dissipation $r = 0.45$ (while fixing other parameters) shown in Fig.\,\ref{fig:abs_peak}(b), the tunneling $\propto t_{h\downarrow}$ becomes irrelevant in a larger regime ($T < 1.1 T_0$), indicating a more strongly suppressed ABS tunneling.
Therefore, we can see that an ABS peak is more sensitive to dissipation.

Briefly, in contrast to a Majorana zero-bias peak, ABS zero-bias peaks do not have a universal height, and in most cases do not necessarily display universality in higher temperature~\cite{SupMat}.
Indeed, following Table.\,\ref{tab:Scaling}, generic ABS conductance curves display universality only when the system flows close enough to the low temperature fixed point. However, our results suggest that a stronger dissipation (larger $r$ but smaller than $0.5$) can easily suppress the ABS peak and provide a sharp difference compared with Majorana peak.

\textbf{\em{Detuned ABS and multiple-ABS scenarios.}}
In real experiment, scaling dimensions listed in Table.\,\ref{tab:Scaling}, which are among our central conclusions, should be checked carefully, as (i) the ABS energy might be finite; and (ii) multiple on-resonance ABSs might talk to the lead simultaneously, and (iii) a Majorana might become surrounded by multiple ABSs~\cite{JJinFeng-2016-PRL}.

As the response to the concern (i), we notice that the ABS detuning Hamiltonian
$H_\text{ABS} = \epsilon a^{\dagger} a$
is highly RG relevant from Eq.~\eqref{eq:rg_energy}.
The detuning energy should thus be considered as another possible low-energy cutoff, in addition to temperature $T$ for the zero-bias situation.
Consequently, if $T\gg\epsilon$, RG flow does not see the coupling $\epsilon$, and then
the conductance behavior then coincides with that when $\epsilon = 0$.
In the opposite limit $T\ll\epsilon$ the single-electron tunneling (e.g., $\psi_{\uparrow}^\dagger a$) requires to pay an extra energy, and is thus suppressed.
In both limits, Andreev tunneling dominates at low temperatures, leading to the same temperature dependence as shown in Table.\,\ref{tab:Scaling}.
Finally, the crossover $T\sim \epsilon$ is not within the universality class, and conductance should not follow any temperature power-law.
The concern (ii), i.e., the multi-ABS scenario, can be analyzed following the g-theorem, which states that the system prefers the lower-entropy ground state \cite{AffleckLudwigPRL91}. ABSs either decouple or become fully hybridized by the lead at low enough temperatures. Of this situation, charge transport is only possible via Andreev reflection. The result is once again the same as in Table.\,\ref{tab:Scaling}.


The situation becomes most interesting if the superconductor hosts a single MZM (either topological or quasi) and one or multiple generic ABSs. At lowest energies, the system behavior is not hard to speculate following our analysis above on the multi-ABS scenario: all ABSs become decoupled or hybridized, and lead-wire charge transport relies on tunneling into the MZM.
Consequently, for generic cases, we expect the same low-temperature behavior as the MZM situation of Table.\,\ref{tab:Scaling}, given low-enough temperatures. However, how the system arrives at this fixed point might depend on, e.g., the relative amplitudes of the lead-ABS and the lead-MZM couplings. Indeed, when lead-ABS coupling is stronger at high temperatures, one might expect the non-trivial transition from ABS-like to MZM-like scaling behaviors as reducing temperature.

Finally, we also notice a related experimental progress in the study of Andreev reflections using a dissipative probe in the SC nanowire devices~\cite{ZhangHaoPre}. 

\textbf{{\em Acknowledgments.}} The work is supported by Natural
Science Foundation of China (Grants No.~11974198 and No.~12004040) and Tsinghua University Initiative Scientific Research Program.

\bibliographystyle{apsrev4-1} 
\bibliography{DissipativeABS,BibFootnotes}

\begin{thebibliography}{81}
\bibitem [{Inc()}]{Incoherence}%
  \BibitemOpen
  \href@noop {} {}\bibinfo {note} {Here we do not break the coherence of two
  sub-processes of a single Andreev tunneling. This is similar as the
  incoherence Andreev reflection in non-equilibrium
  setups~\cite{BezuglyiPRL99,SylviePRL09}, where decoherence only emerges
  between different Andreev reflection processes.}\BibitemShut {Stop}%
\end{thebibliography}

\onecolumngrid
\newpage


\section*{Supplementary material}

# Supplementary Information for "Universal conductance scaling of Andreev reflections using a dissipative probe"

In this supplementary information, we will provide details concerning: (i) Mapping to the Coulomb gas model; (ii) The derivation of RG equations (6) of the main text; (iii) Analysis of the decoherent of an Andreev reflection; (iv) Low-energy features obtained with the effective Hamiltonian; (v) Accidental Majorana Hamiltonian and its mapping to a Sine-Gordon model, and (vi) Plots of ABS conductance curves in the presence of a stronger asymmetry.

## I. MAPPING TO THE COULOMB GAS MODEL

### A. Details of the model

As mentioned in the main text, the system we study can be divided into four parts: the SC nanowire, the lead, the tunneling part, and the dissipative environment. The main target we focus in the SC nanowire is the low-energy state with its most distribution near the tunnel-junction. Such states may arise because of the disorder or the potential variation. In the diagonal form, the Hamiltonian of the nanowire part can be written as $\tilde{H}_{\text{wire}} = \epsilon a^\dagger a + \sum_p \epsilon_p a_p^\dagger a_p + \text{const}$, where $a$ is operator of the state we study. $a_p$ is the operator of the other state that has very little distribution near the junction which are either the widely-spread high-energy quasi-particle state outside the superconducting gap or the state localized away from the tunnel junction. In the BDG formalism, the wavefunction of a state in a superconductor can be described as a four-component vector form $(u_\uparrow, u_\downarrow, v_\uparrow, v_\downarrow)^T$, where $u$ and $v$ are the particle and the hole part, respectively. In our case, the electron operator in the nanowire can be written in terms of quasiparticle and hole operators: $c_\sigma(\tilde{x}) = u_{0,\sigma}(\tilde{x}) a + v_{0,\sigma}^*(\tilde{x}) a^\dagger + \sum_p (u_{p,\sigma}(\tilde{x}) a_p + v_{p,\sigma}^*(\tilde{x}) a_p^\dagger)$ where $\sigma = \uparrow / \downarrow$ labels the spin. $u_{0,\sigma}/v_{0,\sigma}$ and $u_{p,\sigma}/v_{p,\sigma}$ are the particle/hole part of the wave functions of the state we study and other states respectively. Here we use the coordinate $\tilde{x}$ to label the position in the nanowire (note we use $x$ to label the position in the lead). As we have assumed only the state we study has large distribution at the left end of the nanowire. We set the starting point at the left end of the nanowire as $\tilde{x} = 0$, then we have $u_{p,\sigma}(0) \approx 0$, $v_{p,\sigma}(0) \approx 0$. Since the tunnelings are terms like $\psi_\uparrow^\dagger(0) c_\uparrow(0)$ where $\psi_\sigma(x)$ is fermion operator in the lead, the effect of other states can be safely ignored. The four tunneling parameters in the Eq. (2) of the main text have $t_{e\uparrow} \propto u_{0,\uparrow}(0)$, $t_{e\downarrow} \propto u_{0,\downarrow}^*(0)$, $t_{h\uparrow} \propto v_{0,\uparrow}(0)$, $t_{h\downarrow} \propto v_{0,\downarrow}^*(0)$. Because the spin rotation and time reversal symmetries are broken due to the spin-orbit coupling and the magnetic field, the four tunneling parameters can't be reduced.

Next we talk about the details of the bosonization of the semi-infinite lead. The chiral bosonic field is introduced by $\psi_{L,\sigma}(x) = \frac{1}{\sqrt{2\pi\alpha}} F_\sigma e^{-i\Phi_\sigma(-x)}$, $\psi_{R,\sigma}(x) = \frac{1}{\sqrt{2\pi\alpha}} F_\sigma e^{-i\Phi_\sigma(+x)}$ which unfold the right- and left-moving modes onto a single axis extending from $-\infty$ to $+\infty$. It is easy to see that the continuity requirement of $\Phi_\sigma$ at $x = 0$ just coincides with the boundary condition $\psi_{L,\sigma}(x = 0) = \psi_{R,\sigma}(x = 0)$. Then the lead Hamiltonian [Eq.(1) of the main text] becomes

$$H_{\text{lead}} = \frac{v_F}{4\pi} \int_{-\infty}^{+\infty} dx \sum_{\sigma = \uparrow/\downarrow} [\partial_x \Phi_\sigma(x)]^2 . \tag{S1}$$

Then the corresponding action of the lead is

$$S_{\text{lead}} = \frac{v_F}{4\pi} \int dx \int d\tau \sum_\sigma \left[ \frac{1}{v_F^2} (\partial_\tau \Phi_\sigma)^2 + (\partial_x \Phi_\sigma)^2 \right] . \tag{S2}$$

The action is minimized when $\Phi_\sigma(x, \omega_n) = \Phi_\sigma(\omega_n) \exp(-|\omega_n x| / v_F)$ [S1]. $\Phi_\sigma(\omega_n)$ is the Fourier transform of the $\Phi_\sigma$ field at $x = 0$: $\Phi_\sigma(x = 0, \tau) \equiv \Phi_\sigma(\tau) = \frac{1}{\beta} \sum_n \Phi_\sigma(\omega_n) e^{-i\omega_n \tau}$. Integrate out $x$ away from $x = 0$, we have

$$S_{\text{lead}}^{\text{eff}} = \frac{1}{\beta} \sum_{\omega_n} |\omega_n| \sum_\sigma |\Phi_\sigma(\omega_n)|^2 . \tag{S3}$$

### B. Derivation of the Coulomb gas model

The partition function of the system is given by the Eq. (3) of the main text: $Z = \int [D\Phi_\uparrow] [D\Phi_\downarrow] [D\varphi] [Da] e^{-S_{\text{eff}}} e^{-S_{\text{T}}}$, where $S^{\text{eff}}$ is the sum of the actions of the lead and environment: $S^{\text{eff}} = \frac{1}{\beta} \sum_{\omega_n} |\omega_n| \left[ \sum_\sigma |\Phi_\sigma(\omega_n)|^2 + |\varphi(\omega_n)|^2 / 2r \right]$ and $S_{\text{T}}$ is the

action of the tunneling: $S_T = \int_0^\beta d\tau L_T$. The Lagrangian of the tunneling part can be expressed as

$$L_T = -\sqrt{\frac{2}{\pi\alpha}} \left( F_\uparrow e^{i\Phi_\uparrow(x=0)} \; F_\downarrow e^{i\Phi_\downarrow(x=0)} \right) \begin{pmatrix} t_{e\uparrow} & t_{h\uparrow} \\ t_{e\downarrow} & t_{h\downarrow} \end{pmatrix} \begin{pmatrix} a \\ a^\dagger \end{pmatrix} e^{-i\varphi} + c.c. \tag{S4}$$

Expand $Z$ in each powers of $L_T$:

$$Z = \left\langle 1 + \int_0^\beta d\tau_2 \int_0^{\tau_2} d\tau_1 L_T(\tau_2) L_T(\tau_1) + \int_0^\beta d\tau_4 \int_0^{\tau_4} d\tau_3 \int_0^{\tau_3} d\tau_2 \int_0^{\tau_2} d\tau_1 L_T(\tau_4) L_T(\tau_3) L_T(\tau_2) L_T(\tau_1) + \ldots \right\rangle_0, \tag{S5}$$

where we have used $\langle \cdots \rangle_0$ to represent $\int [D\Phi_\uparrow][D\Phi_\downarrow][D\varphi][Da] \cdots$. By grouping the same fields in each term together we can obtain the correlators:

$$\int [Da] \ldots a(\tau_l) a^\dagger(\tau_m) \ldots \times \int [D\varphi] e^{-\frac{1}{\beta}\sum_{\omega_n}|\omega_n|\frac{1}{2r}|\varphi(\omega_n)|^2} \ldots e^{q_e i\varphi(\tau_e)} e^{q_f i\varphi(\tau_f)} \ldots$$

$$\times \int [D\Phi_\uparrow] e^{-\frac{1}{\beta}\sum_{\omega_n}|\omega_n||\phi_\uparrow(\omega_n)|^2} \ldots e^{\alpha_a i\Phi_\uparrow(\tau_a)} e^{\alpha_b i\Phi_\uparrow(\tau_b)} \ldots \times \int [D\Phi_\downarrow] e^{-\frac{1}{\beta}\sum_{\omega_n}|\omega_n||\Phi_\downarrow(\omega_n)|^2} \ldots e^{\eta_c i\Phi_\downarrow(\tau_c)} e^{\eta_d i\Phi_\downarrow(\tau_d)} \ldots$$

$$= \langle \ldots a(\tau_l) a^\dagger(\tau_m) \ldots \rangle \times \left\langle \ldots e^{q_e i\sqrt{2r}\varphi'(\tau_e)} e^{q_f i\sqrt{\frac{2R}{R_Q}}\varphi'(\tau_f)} \ldots \right\rangle$$

$$\times \langle \ldots e^{\alpha_a i\Phi_\uparrow(\tau_a)} e^{\alpha_b i\Phi_\uparrow(\tau_b)} \ldots \rangle \times \langle \ldots e^{\eta_c i\Phi_\downarrow(\tau_c)} e^{\eta_d i\Phi_\downarrow(\tau_d)} \ldots \rangle, \tag{S6}$$

where we have defined $\varphi' = \varphi/\sqrt{2r}$. By evaluating each correlators, we obtain

$$Z = \sum_{\nu=\pm 1} \sum_n \sum_{\{q_i\}} \sum_{\{\alpha_i\}} \sum_{\{\eta_i\}} C_l \int_0^\beta d\tau_{2n} \ldots \int_0^{\tau_3} d\tau_2 \int_0^{\tau_2} d\tau_1 e^{\sum_{i>j} V_{ij}} e^{\nu\epsilon\left[\frac{\beta}{2} + \sum_i (-1)^i \tau_i\right]}, \tag{S7}$$

where $V_{ij} = (2rq_iq_j + \alpha_i\alpha_j + \eta_i\eta_j)\ln(\tau_i - \tau_j)$, $C_t$ is the product of tunneling parameters, which is determined by the series of tunnelings and therefore will depend on $\{q_i, r_i, \alpha_i, \eta_i\}$. Here we have omitted the Klein factors and the constant in the front because they are RG irrelevant.

Because at $\tau_i$, either spin-$\uparrow$ or spin-$\downarrow$ takes part in the tunneling process, only one of $\alpha$ and $\eta$ is non-zero ($\alpha_i = \pm 1$, $\eta_i = 0$ or $\alpha_i = 0$, $\eta_i = \pm 1$). Then we can define $s_i = \alpha_i - \eta_i = \pm 1$ which represents the pure spin injected into the nanowire. And we can also define

$$r_i = \begin{cases} +1 & \text{, after } a^\dagger(\tau_i) \\ -1 & \text{, after } a(\tau_i) \end{cases} \tag{S8}$$

Note each combination of values of $q_i, s_i, r_i$ can define a tunneling process. At $\tau_i$, it can be

| $q_i$ | $r_i$ | $s_i$ | Process | Tunneling parameter | Original parameters |
|---|---|---|---|---|---|
| +1 | +1 | +1 | $\underrightarrow{e \uparrow}$ | $t_{e\uparrow}^*$ | $\alpha_i = +1,\ \eta_i = 0$ |
| | | -1 | $\underrightarrow{e \downarrow}$ | $t_{e\downarrow}$ | $\alpha_i = 0,\ \eta_i = +1$ |
| | -1 | +1 | $\overrightarrow{h \uparrow}$ | $t_{h\uparrow}$ | $\alpha_i = +1,\ \eta_i = 0$ |
| | | -1 | $\overleftarrow{h \downarrow}$ | $t_{h\downarrow}^*$ | $\alpha_i = 0,\ \eta_i = +1$ |
| -1 | +1 | +1 | $\underrightarrow{h \downarrow}$ | $t_{h\downarrow}$ | $\alpha_i = 0,\ \eta_i = -1$ |
| | | -1 | $\overrightarrow{h \uparrow}$ | $t_{h\uparrow}^*$ | $\alpha_i = -1,\ \eta_i = 0$ |
| | -1 | +1 | $\overleftarrow{e \downarrow}$ | $t_{e\downarrow}^*$ | $\alpha_i = 0,\ \eta_i = -1$ |
| | | -1 | $\overleftarrow{e \uparrow}$ | $t_{e\uparrow}$ | $\alpha_i = -1,\ \eta_i = 0$ |

where the process, for example, $\underrightarrow{e \uparrow}$ means an electron with spin up hops into the nanowire, $\overleftarrow{h \downarrow}$ means an electron with spin down is reflected from the nanowire. Physically, $q_i, s_i$ represent the transfer of the electric charge and the spin respectively. In a tunneling process, either an electric charge is injected into ($q_i = +1$) or reflected out of ($q_i = -1$) the SC nanowire. If an up spin is injected into (or a down spin is reflected out of) the nanowire, $s_i = +1$. And if a down spin is injected into (or an up spin is reflected out of) the nanowire, $s_i = -1$. $r_i = \pm 1$ represents the of the excitation of the ABS state as shown in Eq. (S8).

Next we will show that this tunneling model can be mapped to a one-dimensional Coulomb gas model composed of interacting charges. $q_i, s_i, r_i = \pm 1$ can be seen as three types of charges which are located at the site $\tau_i$ of the timeline. From the table above we can conclude that

$$\alpha_i \alpha_j = \frac{1}{2} \left( q_i q_j + s_i s_j \right), \eta_i \eta_j = 0 \text{ or } \alpha_i \alpha_j = 0, \eta_i \eta_j = \frac{1}{2} \left( q_i q_j + s_i s_j \right). \tag{S9}$$

and

$$V_{ij} = \frac{1}{2g} \left( q_i q_j + g s_i s_j \right) \ln \left( \tau_i - \tau_j \right), \tag{S10}$$

which denotes the interaction between charges. $g = \left( 1 + 4Re^2/h \right)^{-1} = (1 + 4r)^{-1}$. Here interactions only exist between same charges ($q_i$s and $s_i$s). Later we will show that the interactions involving $r_i$ and the interactions between different types of charges will appear during the renormalization. Besides, there are two constraints on the charges: first, because in each term of $Z$, the state must evolve back to the starting state, all kinds of charges are required to be neutral $\sum_i^{2n} q_i = 0$, $\sum_i^{2n} s_i = 0$, $\sum_i^{2n} r_i = 0$. Second, because only the terms with $\langle \ldots aa^\dagger aa^\dagger aa^\dagger \ldots \rangle$ are non-zero, we have the constraint that $r_i$ charge must alternate in time $r_i r_{i+1} = -1$. We have shown that in the Table above that each combination of $q$, $r_i$ and $s_i$ can determine a tunneling process. Then in $n$-th term of Eq. (S5), with $i$ from 1 to $2n$, a series of tunnelings is formed. It can be concluded from the table that the total tunneling parameters can be expressed as:

$$C_\mathrm{t} = \prod_{i=1}^{2n} \prod_{\xi} |t_\xi|^{\frac{1}{4}(1 + \lambda_{\xi,1} q_i r_i + \lambda_{\xi,2} q_i s_i + \lambda_{\xi,3} r_i s_i)} \left( \frac{t_\xi}{t_\xi^*} \right)^{-\frac{1}{8} q_i r_i s_i}, \tag{S11}$$

where $\xi$ denotes $e\uparrow, h\uparrow, e\downarrow, h\downarrow$, and $(\lambda_{\xi,1}, \lambda_{\xi,2}, \lambda_{\xi,3})$ can take $(+1, +1, +1), (-1, +1, -1), (+1, -1, -1), (-1, -1, +1)$ respectively. At last, the partition function in Eq. (S7) can be written in the form:

$$Z = \sum_{\nu = \pm 1} \sum_n \sum_{\{q_i\}} \sum_{\{s_i\}} C_\mathrm{t} \int_0^\beta d\tau_{2n} \int_0^{\tau_{2n}} d\tau_{2n-1} \ldots \int_0^{\tau_3} d\tau_2 \int_0^{\tau_2} d\tau_1 e^{\sum_{i>j} V_{ij}} e^{\nu \epsilon \left[ \frac{\beta}{2} + \sum_i (-1)^i \tau_i \right]}. \tag{S12}$$

For a zero-energy state ($\epsilon = 0$ which is the main object we study), Eq. (S12) is exactly a multi-charge Coulomb gas model. Next we will write the interaction as

$$V_{ij} = \frac{1}{2g} \left[ q_i q_j + K_1 \left( q_i r_j + r_i q_j \right) + g s_i s_j + K_2 g \left( s_i r_j + r_i s_j \right) + K_3 r_i r_j \right] \ln \left( \frac{\tau_i - \tau_j}{\tau_c} \right), \tag{S13}$$

where $\tau_c$ is a short-time cutoff which gives the minimum distance between two sites $|\tau_i - \tau_j| \geq \tau_c, (i \neq j)$. $1/\tau_c$ refers to the high-energy cutoff of the system. Next we will do the RG by changing $\tau_c$. Here we also introduce another three interactions with strength $K_1, K_2, K_3$. During the RG process, $K_1, K_2, K_3$ will flow away from the original value zero.

## II. DERIVATION OF THE RG EQUATIONS

The RG process is simmilar to that of [S1, S2] consisting of the following steps: first by integrating all the charge pairs (dipoles) whose distance is between $\tau_c$ and $\tau_c + d\tau_c$, the cutoff $\tau_c$ can be enlarged to $\tau_c + d\tau_c$; then rescale $\tau$ to a bigger $\tau_\mathrm{new}$ with $\tau_\mathrm{new} = \tau \left( \tau_c + d\tau_c \right) / \tau_c$.

In the step of enlarging $\tau_c$, the most influential factor is the neutral charge pair (dipole) whose distance is between $\tau_c$ and $\tau_c + d\tau_c$. The partition function can be written as the sum of no such charge pairs and one charge pair. The terms with more charges between $\tau_c$ and $\tau_c + d\tau_c$ can be omitted because they are small values of higher order. Then

$$Z = \sum_n \sum_{\{q_i\}} \sum_{\{r_i\}} \sum_{\{s_i\}} C_\mathrm{t} \int_0^\beta d\tau_{2n} \int_0^{\tau_{2n} - \tau_c - d\tau_c} d\tau_{2n-1} \ldots \int_0^{\tau_3 - \tau_c - d\tau_c} d\tau_2 \int_0^{\tau_2 - \tau_c - d\tau_c} d\tau_1 \exp \left( \sum_{i>j} V_{ij} \right)$$
$$\times \left\{ 1 + \sum_{\text{all dipoles}} C_\mathrm{t} \left( q', r', s', q'', r'', s'' \right) \sum_i \int_{2\tau_c + \tau_i}^{\tau_{i+1} - \tau_c} d\tau' \int_{\tau' - \tau_c - d\tau_c}^{\tau' - \tau_c} d\tau'' \exp \left[ V \left( \tau', \tau'' \mid \{\tau_j\} \right) \right] \right\}. \tag{S14}$$

Here the position of the charge pair is labeled by $\tau'$ and $\tau''$ and their distance $\tau_c < \tau' - \tau'' < \tau_c + d\tau_c$. When the pair is between $\tau_i < \tau'' < \tau' < \tau_{i+1}$, we can write value range of $\tau'$ and $\tau''$, as shown by the the upper and lower bound of their

integrals above. In Eq. (S14), $C_t\left(q', r', s', q'', r'', s''\right)$ is the product of tunneling parameters at $\tau'$ and $\tau''$, $V\left(\tau', \tau'' \mid \{\tau_j\}\right)$ is the interaction between the dipole and all the other charges. After some algebra, it can be written as

$$V\left(\tau', \tau'' \mid \{\tau_j\}\right) = \sum_j \frac{1}{2g}\left[q'q_j + gs's_j + K_1\left(q'r_j + r'q_j\right) + K_2 g\left(s'r_j + r's_j\right) + K_3 r'r_j\right]\tau_c\partial_{\tau'}\ln\frac{\tau' - \tau_j}{\tau_c} \quad \text{(S15)}$$

Here we consider a zero-energy ABS $\epsilon = 0$. We will extend our result to the case of $\epsilon \neq 0$ at last. Because of the constraint $r_i r_{i+1} = -1$ (note $\tau_i < \tau'' < \tau' < \tau_{i+1}$), we can have $r' = r_i$. For the neutral pairs ($q' + q'' = 0$, $r' + r'' = 0$ and $s' + s'' = 0$, the allowed processes (dipoles) are

| process at $\tau'', \tau'$ | $e\uparrow e\uparrow$ | $e\uparrow e\uparrow$ | $e\downarrow e\downarrow$ | $e\downarrow e\downarrow$ | $h\uparrow h\uparrow$ | $h\uparrow h\uparrow$ | $h\downarrow h\downarrow$ | $h\downarrow h\downarrow$ |
|---|---|---|---|---|---|---|---|---|
| $r_i$ | -1 | +1 | -1 | +1 | -1 | +1 | -1 | +1 |
| $q'$ | -1 | +1 | -1 | +1 | +1 | -1 | +1 | -1 |
| $s'$ | -1 | +1 | +1 | -1 | +1 | -1 | -1 | +1 |
| relation | $q' = r_i, s' = r_i$ | | $q' = r_i, s' = -r_i$ | | $q' = -r_i, s' = -r_i$ | | $q' = -r_i, s' = r_i$ | |
| $C_t\left(q', r', s', q'', r'', s''\right)$ | $|t_{e\uparrow}|^2$ | | $|t_{e\downarrow}|^2$ | | $|t_{h\uparrow}|^2$ | | $|t_{h\downarrow}|^2$ | |

The fifth row of the table shows us the relation of the charges at $\tau'$ and $\tau_i$. The sixth row shows the tunneling parameters of each dipole. Using these relations, Eq. (S14) can be further transformed to

$$Z = \sum_n \sum_{\{q_i\}}\sum_{\{r_i\}}\sum_{\{s_i\}} C_t \int_0^\beta d\tau_{2n}\int_0^{\tau_{2n} - \tau_c - d\tau_c} d\tau_{2n-1}\ldots\int_0^{\tau_3 - \tau_c - d\tau_c} d\tau_2 \int_0^{\tau_2 - \tau_c - d\tau_c} d\tau_1 \exp\left(\sum_{i > j}V_{ij}\right)$$
$$\times \exp\left(-2\tau_c d\tau_c \sum_{i \neq j}\frac{1}{2g}U_{ij}\ln\frac{\tau_i - \tau_j}{\tau_c}\right), \quad \text{(S16)}$$

where

$$\begin{aligned}U_{ij} =&\, C_t\left(q', r', s', q'', r'', s''\right)\left[q'q_j + gs's_j + K_1\left(q'r_j + r'q_j\right) + K_2 g\left(s'r_j + r's_j\right) + K_3 r'r_j\right]\\
=&\left[\left(|t_{e\uparrow}|^2 - |t_{h\uparrow}|^2 + |t_{e\downarrow}|^2 - |t_{h\downarrow}|^2\right) + \left(|t_{e\uparrow}|^2 + |t_{h\uparrow}|^2 + |t_{e\downarrow}|^2 + |t_{h\downarrow}|^2\right)K_1\right]r_iq_j\\
&+ g\left[\left(|t_{e\uparrow}|^2 - |t_{h\uparrow}|^2 - |t_{e\downarrow}|^2 + |t_{h\downarrow}|^2\right) + \left(|t_{e\uparrow}|^2 + |t_{h\uparrow}|^2 + |t_{e\downarrow}|^2 + |t_{h\downarrow}|^2\right)K_2\right]r_is_j\\
&+ \left[\left(|t_{e\uparrow}|^2 - |t_{h\uparrow}|^2 + |t_{e\downarrow}|^2 - |t_{h\downarrow}|^2\right)K_1 + \left(|t_{e\uparrow}|^2 - |t_{h\uparrow}|^2 - |t_{e\downarrow}|^2 + |t_{h\downarrow}|^2\right)K_2 g\right.\\
&\left.+ \left(|t_{e\uparrow}|^2 + |t_{h\uparrow}|^2 + |t_{e\downarrow}|^2 + |t_{h\downarrow}|^2\right)K_3\right]r_ir_j\end{aligned} \quad \text{(S17)}$$

Compare Eq. (S16) with the original partition function Eq.(S12), we can see only $K_1$, $K_2$ and $K_3$ are slightly shifted. Their flow equations are:

$$\frac{dK_1}{d\ln\tau_c} = -2\tau_c^2\left[\left(|t_{e\uparrow}|^2 - |t_{h\uparrow}|^2 + |t_{e\downarrow}|^2 - |t_{h\downarrow}|^2\right) + \left(|t_{e\uparrow}|^2 + |t_{h\uparrow}|^2 + |t_{e\downarrow}|^2 + |t_{h\downarrow}|^2\right)K_1\right], \quad \text{(S18a)}$$

$$\frac{dK_2}{d\ln\tau_c} = -2\tau_c^2\left[\left(|t_{e\uparrow}|^2 - |t_{h\uparrow}|^2 - |t_{e\downarrow}|^2 + |t_{h\downarrow}|^2\right) + \left(|t_{e\uparrow}|^2 + |t_{h\uparrow}|^2 + |t_{e\downarrow}|^2 + |t_{h\downarrow}|^2\right)K_2\right], \quad \text{(S18b)}$$

$$\begin{aligned}\frac{dK_3}{d\ln\tau_c} = &-4\tau_c^2\left[\left(|t_{e\uparrow}|^2 - |t_{h\uparrow}|^2 + |t_{e\downarrow}|^2 - |t_{h\downarrow}|^2\right)K_1 + g\left(|t_{e\uparrow}|^2 - |t_{h\uparrow}|^2 - |t_{e\downarrow}|^2 + |t_{h\downarrow}|^2\right)K_2\right.\\
&\left.+ \left(|t_{e\uparrow}|^2 + |t_{h\uparrow}|^2 + |t_{e\downarrow}|^2 + |t_{h\downarrow}|^2\right)K_3\right], \quad \text{(S18c)}\end{aligned}$$

where we have used $\sum_{i \neq j}r_ir_j = 2\sum_{i > j}r_ir_j$, $\sum_{i \neq j}r_iq_j = \sum_{i > j}\left(r_iq_j + q_ir_j\right)$ and $\sum_{i \neq j}r_is_j = \sum_{i > j}\left(r_is_j + s_ir_j\right)$. From Eq. (S18a-S18c) above, we can easily have

$$\frac{d}{d\ln\tau_c}\left(K_1^2 + gK_2^2 - K_3\right) = -4\tau_c^2\left(|t_{e\uparrow}|^2 + |t_{h\uparrow}|^2 + |t_{e\downarrow}|^2 + |t_{h\downarrow}|^2\right)\left(K_1^2 + gK_2^2 - K_3\right) \quad \text{(S19)}$$

which shows that $K_1^2 + gK_2^2 - K_3$ will always flows to 0. Then $K_3$ is not independent and $K_3 = K_1^2 + gK_2^2$. Therefore, the above three RG equations can be reduced to two equations of $K_1$ and $K_2$.

Next, we rescale $\tau$ to make partition function back to the original form. We need a large $\tau_{\text{new}}$ with $\tau_{\text{new}} = \tau \left( \tau_c + d\tau_c \right) / \tau_c$ to make the average charges per unit $\tau_{\text{new}}$ in the partition function Eq. (S16) same as the average charges per unit $\tau$ in the partition function Eq. (S12). When using $\tau_{\text{new}}$, a factor will be produced in Eq. (S16):

$$\left( \frac{\tau_c}{\tau_c + d\tau_c} \right)^{2n\left[ 1 - \frac{1}{4g}\left( 1 + g + K_1^2 + gK_2^2 \right) \right] - \frac{1}{2g}\left( K_1 \sum_i q_i r_i + gK_2 \sum_i s_i r_i \right)},$$ 

(S20)

which comes from the $d\tau_i$ and the interaction $\ln\left[ \left( \tau_i - \tau_j \right) / \tau_c \right]$ in Eq. (S16). To make the partition back to the original form, this factor must be absorbed by the tunneling parameters, which slightly change each tunneling parameter $t_\xi$. The resulting flow equations are

$$\frac{dt_\xi}{d\ln\tau_c} = \left[ 1 - \frac{\left( K_1 + \delta_{\xi,1} \right)^2 + g\left( K_2 + \delta_{\xi,2} \right)^2}{4g} \right] t_\xi,$$

(S21)

where $(\delta_{\xi,1}, \delta_{\xi,2})$ equals $(+1, +1)$, $(-1, -1)$, $(+1, -1)$, $(-1, +1)$ for $\xi = e\uparrow, h\uparrow, e\downarrow$ and $h\downarrow$ respectively.

At last we study the case when the ABS has finite energy $\epsilon \neq 0$. During the renormalization process, there will be no difference when enlarging the cutoff $\tau_c$. However, During the rescaling by $\tau_i \rightarrow \tau_i\tau_c / \left( \tau_c + d\tau_c \right)$, the energy of the ABS shall be renormalized as $\epsilon \rightarrow \epsilon \left( \tau_c + d\tau_c \right) / \tau_c$. This will leave us with the last RG equation:

$$\frac{d\epsilon}{d\ln\tau_c} = \epsilon.$$

(S22)

Together Eq. (S18a,S18b,S21,S22) are the total RG equations shown in the Eq.(6) of the main text.

## III. INCOHERENCE OF TUNNELINGS

In this section, we explain the source of incoherence in an Andreev tunneling. Without the loss of generality, we consider the opposite-spin Andreev reflection $\mathcal{O}_{\text{AR}} = \psi_\uparrow^\dagger a \exp(-i\varphi)\psi_\downarrow^\dagger a^\dagger \exp(-i\varphi) + h.c.$ that consists of two sub-processes $\psi_\uparrow^\dagger a \exp(-i\varphi)$ and $\psi_\downarrow^\dagger a^\dagger \exp(-i\varphi)$.

In conformal field theory, the scaling dimension $\zeta_\mathcal{O}$ of an operator $\mathcal{O}$ is determined by its long-time correlation

$$\langle \mathcal{O}(t)\mathcal{O}(0) \rangle = \frac{\mathcal{C}_\mathcal{O}}{t^{2\zeta_\mathcal{O}}},$$

(S23)

where $\mathcal{C}_\mathcal{O}$ is a non-universal normalization constant. For perfect systems, dissipative phases have the same time argument and can be combined. The scaling dimension of $\mathcal{O}_{\text{AR}}$ is simply calculated via

$$\langle \mathcal{O}_{\text{AR}}(t)\mathcal{O}_{\text{AR}}(0) \rangle_{\text{coherent}} = \langle H_{\text{T}}^0(t)H_{\text{T}}^0(0) \rangle \langle e^{-2i\varphi(t)}e^{2i\varphi(0)} \rangle$$
$$\propto \langle H_{\text{T}}^0(t)H_{\text{T}}^0(0) \rangle \frac{1}{(\omega_R t)^{8r}},$$

(S24)

where $H_{\text{T}}^0 = \psi_\uparrow^\dagger a\psi_\downarrow^\dagger a^\dagger + h.c.$ refers to the dissipation-free tunneling Hamiltonian, and $1/\omega_R$ refers to the short-time cutoff. The coherent correlation Eq. (S24) has a long-time tail, when $t \gg 1/\omega_R$.

On the contrary, for an imperfect system, the time of relaxation is influenced by the transient states in the superconducting lead. The time argument of $\varphi$ thus becomes undetermined. The correlation among dissipative phase then becomes

$$\iint d\delta t\, d\delta t'\, \mathcal{P}(\delta t, \delta t') \left\langle e^{-i\varphi(t+\delta t)}e^{-i\varphi(t+\delta t')}e^{i\varphi(\delta t)}e^{i\varphi(\delta t')} \right\rangle$$
$$= \iint d\delta t\, d\delta t'\, \mathcal{P}(\delta t, \delta t') \left[ \frac{(\delta t - \delta t' - i/\omega_R)^2}{(t - i/\omega_R)(t + \delta t - \delta t' - i/\omega_R)(t + \delta t' - \delta t - i/\omega_R)(t - i/\omega_R)} \right]^{2r}$$
$$= \iint d\delta t\, d\delta t'\, \mathcal{P}(\delta t, \delta t') \left[ \frac{(\Delta t - i/\omega_R)^2}{(t - i/\omega_R)(t + \Delta t - i/\omega_R)(t - \Delta t - i/\omega_R)(t - i/\omega_R)} \right]^{2r}$$
$$= \iint d\delta t\, d\delta t'\, \mathcal{P}(\delta t, \delta t')\, \mathcal{C}_{\text{diss}}(t, \Delta t, \omega_R),$$

(S25)

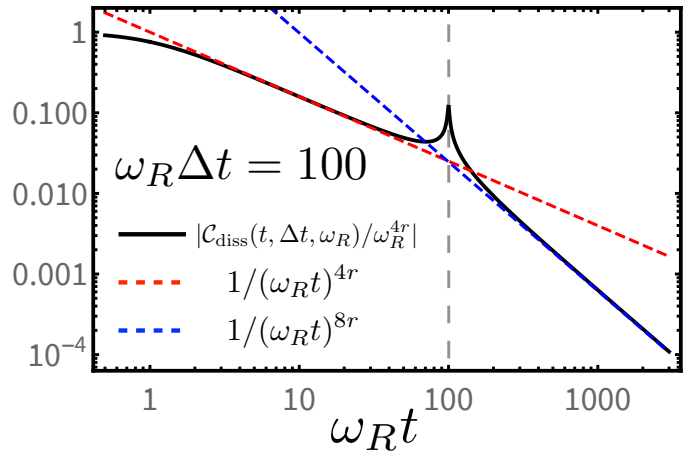

Figure S1. The correlation function Eq. (S26), when $\Delta t = 100/\omega$. The correlation changes from the short-time $\sim 1/(\omega_R t)^{4r}$ to long-time behavior $\sim 1/(\omega_R t)^{8r}$, around $t \sim \Delta t$.

where $\mathcal{P}(\delta t, \delta t') > 0$ refers to the probability of the delay time $\delta t$ and $\delta t'$ for two dissipative phases, and $\Delta t = \delta t - \delta t'$ is their difference. For a better understanding of Eq. (S25), we plot the correlation function of the dissipative fields

$$\mathcal{C}_{\text{diss}}(t, \Delta t, \omega_R) = \left[ \frac{(\Delta t - i/\omega_R)^2}{(t - i/\omega_R)(t + \Delta t - i/\omega_R)(t - \Delta t - i/\omega_R)(t - i/\omega_R)} \right]^{2r} \tag{S26}$$

in Fig. S1. Two major pieces of information are conveyed. As the first one, correlation function displays a polynomial decaying pattern only when $t > 1/\omega_R$. Indeed, $\omega_R \gg T$ is known as the prerequisite to observe the power-law feature in $T$ [S3]. More importantly, the long-time correlation changes from $(\omega_R t)^{4r}$ when $t \ll \Delta t$, to $(\omega_R t)^{8r}$ in the opposite regime when $t \gg \Delta t$. Which power-law is observed experimentally then depends on the relative amplitudes of the transition value $\Delta t$, and the energy-cutoff inverse, e.g., $1/T$.

For simplicity, we assume that the distribution $\mathcal{P}(\delta t, \delta t') = \mathcal{P}(\Delta t)$ depends on the time-difference $\Delta t$, and is mostly distributed around $\Delta t \sim t_{\text{imp}}$. Following the analysis above, we expect different conductance features under different temperatures. When the temperature $T \gg 1/t_{\text{imp}}$, the decreasing of correlation in time terminates at $t \sim 1/T \ll t_{\text{imp}}$. The correlation $\sim 1/(\omega_R t)^{4r}$, leading to the zero-bias conductance feature $G \propto T^{4r}$; In contrast, when $T \ll 1/t_{\text{imp}}$, the longer-time decreasing pattern changes to $\sim 1/(\omega_R t)^{8r}$, and the conductance feature instead becomes $G \propto T^{8r}$.

In real experiments, the incoherence between two sub-processes of an Andreev reflection can also occur if the temperature is not low enough, where the ABS has not been strongly hybridized by the lead. Of this situation, the change of the ABS status can not be considered as a virtual process, thus reducing the coherence between two sub-processes. However, in this situation zero-bias conductance is expected as non-universal.

## IV. LOW-ENERGY FEATURES OBTAINED WITH EFFECTIVE HAMILTONIAN

In the main text, we show that the tunneling into a dissipative ABS can be studied with Coulomb gas RG method, through which we obtain different conductance power-laws near different fixed points. In this section, we show that these conductance features can be obtained also by analysing the effective Hamiltonian. For simplicity, in this section we assume all tunneling parameters as real numbers. Indeed, as shown in the main text, phases of them will not enter RG equations, where only their absolute squares matter.

## A. The dissipation-free situation

We begin with the the dissipation-free tunneling Hamiltonian in another form

$$
\begin{aligned}
H_T &= (t_\uparrow + \delta t_\uparrow)\psi_\uparrow^\dagger a + (t_\uparrow - \delta t_\uparrow)\psi_\uparrow^\dagger a^\dagger + (t_\downarrow + \delta t_\downarrow)\psi_\downarrow^\dagger a + (t_\downarrow - \delta t_\downarrow)\psi_\downarrow^\dagger a^\dagger \\
&\quad - (t_\uparrow + \delta t_\uparrow)\psi_\uparrow a^\dagger - (t_\uparrow - \delta t_\uparrow)\psi_\uparrow a - (t_\downarrow + \delta t_\downarrow)\psi_\downarrow a^\dagger - (t_\downarrow - \delta t_\downarrow)\psi_\downarrow a \\
&= \left[t_\uparrow(\psi_\uparrow^\dagger - \psi_\uparrow) + t_\downarrow(\psi_\downarrow^\dagger - \psi_\downarrow)\right](a + a^\dagger) + \left[\delta t_\uparrow(\psi_\uparrow^\dagger + \psi_\uparrow) + \delta t_\downarrow(\psi_\downarrow^\dagger + \psi_\downarrow)\right](a - a^\dagger),
\end{aligned}
\tag{S27}
$$

where $t_\sigma = (t_{e\sigma} + t_{h\sigma})/2$ and $\delta t_\sigma = (t_{e\sigma} - t_{h\sigma})/2$, respectively, refer to the coupling average and the asymmetry of spin $\sigma$. We interpret $\delta t_\sigma$ as asymmetry, since the lead-ABS coupling becomes equivalent to the lead-MZM coupling when $\delta t_\sigma = 0$ for both spins. For simplicity, in this section we assume all tunneling amplitudes are real numbers.

Following Eq. (S27), for the dissipation-free situation, two impurity Majorana fermions $(a + a^\dagger)/\sqrt{2}$ and $(a - a^\dagger)/i\sqrt{2}$ couple to two independent lead Majoranas. These two lead-MZM couplings thus flow independently, and both approach perfect at zero temperature, even if the coupling to $(a - a^\dagger)/i\sqrt{2}$ is negligible $\delta t_\sigma \ll t_\sigma$ initially.

To better see how asymmetry influences the conductance, we define the lead operators after the rotation

$$
\psi_{\uparrow'}^\dagger = \frac{t_\uparrow \psi_\uparrow^\dagger + t_\downarrow \psi_\downarrow^\dagger}{\sqrt{t_\uparrow^2 + t_\downarrow^2}}, \quad \psi_{\downarrow'}^\dagger = \frac{t_\downarrow \psi_\uparrow^\dagger - t_\uparrow \psi_\downarrow^\dagger}{\sqrt{t_\uparrow^2 + t_\downarrow^2}},
\tag{S28}
$$

where $\psi_{\uparrow'}$ and $\psi_{\downarrow'}$ refer to fermions with spin directions defined by the relative amplitudes of $t_\uparrow$ and $t_\downarrow$. For instance, they have spin up and down along the $x$-axis, when $t_\uparrow = t_\downarrow$.

With fermions of Eq. (S28), the tunneling Hamiltonian becomes

$$
\begin{aligned}
H_T &= \sqrt{t_\uparrow^2 + t_\downarrow^2}(\psi_{\uparrow'}^\dagger - \psi_{\uparrow'})(a + a^\dagger) \\
&\quad + \frac{1}{\sqrt{t_\uparrow^2 + t_\downarrow^2}}\left[(\delta t_\uparrow t_\uparrow + \delta t_\downarrow t_\downarrow)(\psi_{\uparrow'}^\dagger + \psi_{\uparrow'}) + (\delta t_\uparrow t_\downarrow - \delta t_\downarrow t_\uparrow)(\psi_{\downarrow'}^\dagger + \psi_{\downarrow'})\right](a - a^\dagger),
\end{aligned}
\tag{S29}
$$

where the first line drives the system towards the critical point with perfect conductance, for the symmetric situation $\delta t_\uparrow = \delta t_\downarrow = 0$.

Operators of the second line, on the other hand, have different influences on the conductance. As the starter, the second spin $\psi_{\downarrow'}$ decouples if $\delta t_\uparrow t_\downarrow = \delta t_\downarrow t_\uparrow$, where the system Hamiltonian becomes effectively spinless, and equivalent to that of the dissipative resonant level with $r = 1$ [S4] if $\delta t_\uparrow t_\uparrow + \delta t_\downarrow t_\downarrow$ is finite. Following knowledge of a dissipative resonant level from, e.g., Ref. [S4], the asymmetry in $\psi_{\uparrow'}$ drives the system towards the Fermi-liquid-like fixed point that has a vanishing current. Indeed, when asymmetry grows to have the same amplitude as the symmetric one, the lead-ABS tunneling becomes equivalent to $\psi_{\uparrow'}^\dagger a + a^\dagger \psi_{\uparrow'}$, through which the ABS is completely hybridized at zero temperature. The lead-superconductor communication is then turned off, leading to a vanishing conductance.

The situation becomes distinct if $\delta t_\uparrow t_\downarrow \neq \delta t_\downarrow t_\uparrow$ and $\delta t_\uparrow t_\uparrow + \delta t_\downarrow t_\downarrow = 0$, where instead $(\psi_{\uparrow'} + \psi_{\uparrow'}^\dagger)/\sqrt{2}$ decouples. Of this situation, impurity-MZMs $(a + a^\dagger)/\sqrt{2}$ and $(a - a^\dagger)/i\sqrt{2}$ couples to different spin channels. Both lead-MZM couplings flow independently, leading to the perfect conductance $4e^2/h$ at zero temperature.

Finally, we move to the most general situation where two asymmetries of Eq. (S29) coexist. Of this situation $(\psi_{\uparrow'} + \psi_{\uparrow'}^\dagger)/\sqrt{2}$ and $(\psi_{\downarrow'} + \psi_{\downarrow'}^\dagger)/\sqrt{2}$ co-operate in the coupling to the MZM $(a - a^\dagger)/i\sqrt{2}$. The equilibrium zero-temperature conductance then becomes

$$
G(T = 0) = \frac{2e^2}{h}\frac{(\delta t_\uparrow t_\downarrow - \delta t_\downarrow t_\uparrow)^2}{(\delta t_\uparrow^2 t_\downarrow^2 + \delta t_\downarrow^2 t_\uparrow^2)}.
\tag{S30}
$$

In contrast to the dissipative situation analyzed in the main text, generically the dissipation-free situation has a non-universal finite conductance plateau at zero temperature. This zero-temperature finite conductance is experimentally less welcome, as the conductance plateau makes it much harder to witness the conductance power-law at low temperatures. More importantly, this finite value is exactly the reason that the MZM measurement suffers from ABS interruptions.

## B. Dissipative situation

Now we move to the dissipative situation, where the tunneling instead becomes

$$
\begin{aligned}
H_{\mathrm{T}} = {}& \sqrt{t_\uparrow^2 + t_\downarrow^2} \left( \psi_{\uparrow\prime}^\dagger e^{-i\varphi} - \psi_{\uparrow\prime} e^{i\varphi} \right)(a + a^\dagger) \\
& + \frac{1}{\sqrt{t_\uparrow^2 + t_\downarrow^2}} \left[ (\delta t_\uparrow t_\uparrow + \delta t_\downarrow t_\downarrow)(\psi_{\uparrow\prime}^\dagger e^{-i\varphi} + \psi_{\uparrow\prime} e^{i\varphi}) + (\delta t_\uparrow t_\downarrow - \delta t_\downarrow t_\uparrow)(\psi_{\downarrow\prime}^\dagger e^{-i\varphi} + \psi_{\downarrow\prime} e^{i\varphi}) \right](a - a^\dagger),
\end{aligned}
\tag{S31}
$$

where $\varphi$ is the dissipative phase defined in the main text. In contrast to the DRLM situation, this phase can not be simply combined with lead fields, and are thus much more complicated to deal with. The mapping to the DRLM is thus only possible if $\delta t_\uparrow t_\downarrow - \delta t_\downarrow t_\uparrow = 0$, where the sector $\psi_{\downarrow\prime}$ has decoupled. For this case, we bosonize fermion $\psi_{\uparrow\prime}$

$$
\psi_{\uparrow\prime}^\dagger = \frac{U_{\uparrow\prime}}{\sqrt{2\pi a_0}} e^{-i\Phi_{\uparrow\prime}},
\tag{S32}
$$

where $U_{\uparrow\prime}$ refers to the Klein factor, and $a_0$ is the lattice constant. We then combine lead boson and the dissipative phase $\Phi \equiv (\Phi_{\uparrow\prime} + \varphi)/\sqrt{1 + 2r}$, to write the asymmetric coupling as $(\psi_{\uparrow\prime}^\dagger e^{-i\varphi} + \psi_{\uparrow\prime} e^{i\varphi}) \propto \cos(\sqrt{1 + 2r}\Phi)$. In comparison to Ref. [S4] where the detuning is written as $\propto \cos(\sqrt{(1 + r_{\mathrm{DRLM}})/2}\,\phi_f')$ (where $\phi_f'$ refers to the charge flavor of a DRLM after being combine with dissipation), our Hamiltonian Eq. (S31) becomes equilvalent to a DRLM with the dissipation $r_{\mathrm{DRLM}} = 1 + 4r$. Based what the knowledge of the DRLM, at low-enough energies, the leading irrelevant tunneling operator has the scaling dimension $2 + 4r$ (it is $1 + r_{\mathrm{DRLM}}$ for the DRLM). This scaling dimension agrees with that from RG equations (6) when $\delta t_\uparrow t_\downarrow - \delta t_\downarrow t_\uparrow = 0$. Indeed, the decoupling of $\psi_{\downarrow\prime}$ forbids the different-spin Andreev reflection. The same-spin Andreev tunneling operator, with the scaling $2 + 4r$, determines the low-temperature conductance feature. This scaling dimension is larger in comparison to that of a opposite-spin Andreev tunneling, and will hardly be observed in real experiments.

Finally, we discuss a special ABS peak that requires fine-tuning $|t_{e\uparrow}|^2 - |t_{h\uparrow}|^2 + |t_{e\downarrow}|^2 - |t_{h\downarrow}|^2 = 0$, and $t_{e\uparrow} t_{h\downarrow} \neq t_{e\downarrow} t_{h\uparrow}$. For simplicity, we take one of its special scenarios, i.e., $t_{e\uparrow} = t_{h\uparrow}$ and $t_{e\downarrow} = -t_{h\downarrow}$ to study the conductance feature. Of this scenario, RG equations (6) behave highly similarly as the accidental MZM situation, since (i) $K_1 = K_2 = 0$, and (ii) all four tunneling parameters have the same scaling dimension even after the RG flow. However, in contrast to the accidental MZM situation, now both impurity-MZMs are involved in the transport, by coupling to lead fermions that belong to different spins. As the impurity entropy equals $\ln\sqrt{2 + 4r} > \ln\sqrt{2}$ of the accidental MZM scenario (under-screened), the couplings to two impurity-MZMs flow independently, and both become perfect at zero temperature. At finite temperatures, the lead-MZM couplings are relaxed, leading to the zero-bias conductance $4e^2/h - G \propto T^{\frac{2-4r}{1+2r}}$.

## V. ACCIDENTAL MZM SITUATION

For the accidental MZM due to fine tuning, the dissipative Hamiltonian Eq. (S31) becomes

$$
H_{\mathrm{MZM}} = t_{\mathrm{MZM}} \left( \psi_{\uparrow\prime}^\dagger e^{-i\varphi} - \psi_{\uparrow\prime} e^{i\varphi} \right)(a + a^\dagger),
\tag{S33}
$$

where $t_{\mathrm{MZM}} = \sqrt{t_\uparrow^2 + t_\downarrow^2}$ refers to the coupling to the MZM. After bosonization (S32) and define the field $\Phi \equiv (\Phi_{\uparrow\prime} + \varphi)/\sqrt{1 + 2r}$, Eq. (S33) becomes

$$
H_{\mathrm{MZM}} = -i t_{\mathrm{MZM}} \sqrt{\frac{2}{\pi a}} \sin(\sqrt{1 + 2r}\Phi)(a + a^\dagger),
\tag{S34}
$$

where the Klein factor has been neglected. Eq. (S34) is the Hamiltonian of a boundary Sine-Gordon model, except for the coupling to the accidental MZM $(a + a^\dagger)/\sqrt{2}$. However, since the rest of the ABS decouples from the system, the accidental MZM has no influence on the transport behavior. Instead, its presence only contributes to a residue entropy $\ln(2 + 4r)$ at zero temperature.

## VI. OTHER ABS PEAKS

In the main text we have discussed an ABS peak that has a significant peak height, and arrives at its peak value at relatively low temperature. This ABS peak occurs in systems with weak asymmetry small weak dissipation. We also show that when

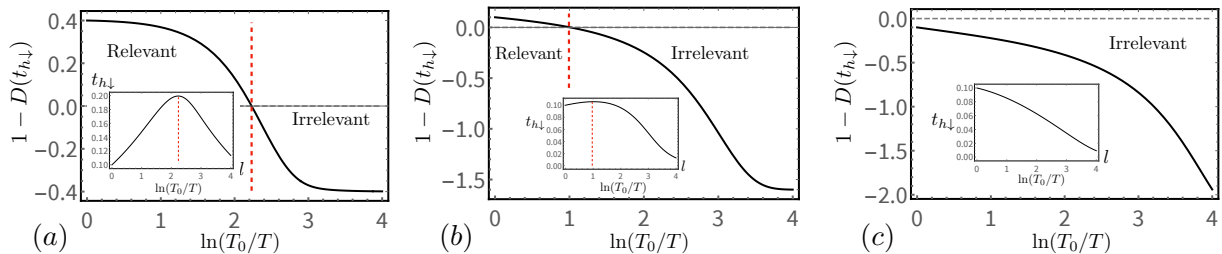

Figure S2. Other ABS peaks that might be observed experimentally. All parameters are the same as Fig. 3 of the main text, expect (a) The value of $t_{h\downarrow}$ decreases, leading to a stronger asymmetry. (b) $r = 0.4$ instead, and (c) $r = 0.6$.

dissipation increases, the peak occurs at a much higher temperature, with a much smaller peak height. In this section we instead show the influence of asymmetry, in Fig. S2.

As the starter, we decrease the value of $t_{h\downarrow}$ in (a), through which a larger asymmetry has been introduced. Clearly, now the ABS conductance peak has a much smaller value. And the system arrives at this peak value at a relatively higher temperature (smaller value of $\ln(T_0/T)$). In (b), we also increase the dissipation amplitude from $r = 0.2$ to $r = 0.4$, while keeping other parameters as that of (a). The conductance is strongly suppressed by both asymmetry and dissipation. Finally, we move to $r = 0.6$ in (c), where the conductance peak completely disappears.



\end{document}